\documentclass[final]{elsarticle}
\usepackage{amsmath,amssymb,amsfonts}
\usepackage{algorithmic}
\usepackage{graphicx}
\usepackage{textcomp}
\usepackage{xcolor}
\usepackage{slashbox}
\usepackage{tabularx}
\usepackage{adjustbox}

\def\BibTeX{{\rm B\kern-.05em{\sc i\kern-.025em b}\kern-.08em
    T\kern-.1667em\lower.7ex\hbox{E}\kern-.125emX}}
\newcolumntype{L}[1]{>{\raggedright\arraybackslash}p{#1}}
\newcolumntype{C}[1]{>{\centering\arraybackslash}p{#1}}
\newcolumntype{R}[1]{>{\raggedleft\arraybackslash}p{#1}}

\begin{document}

\begin{frontmatter}

\title{A Deep Neural Network for Audio Classification with a Classifier Attention Mechanism}


\author[mymainaddress]{Haoye Lu\corref{mycorrespondingauthor}}
\cortext[mycorrespondingauthor]{Corresponding author}
\ead{hlu044@uottawa.ca}

\author[mymainaddress]{Haolong Zhang}
\ead[url]{hzhan006@uottawa.ca}

\author[mymainaddress]{Amit Nayak}
\ead[url]{anaya085@uottawa.ca}

\address[mymainaddress]{University of Ottawa, Ottawa, Ontario, Canada, K1N 6N5}

\begin{abstract}
Audio classification is considered as a challenging problem in pattern recognition. Recently, many algorithms have been proposed using deep neural networks. In this paper, we introduce a new attention-based neural network architecture called Classifier-Attention-Based Convolutional Neural Network (CAB-CNN). The algorithm uses a newly designed architecture consisting of a list of simple classifiers and an attention mechanism as a classifier selector. This design significantly reduces the number of parameters required by the classifiers and thus their complexities. In this way, it becomes easier to train the classifiers and achieve a high and steady performance. Our claims are corroborated by the experimental results. Compared to the state-of-the-art algorithms, our algorithm achieves more than $10\%$ improvements on all selected test scores.
\end{abstract}


\end{frontmatter}


\section{Introduction}
\label{sec:intro}
Sounds contain rich information and help people sense the environments around them. People are able to recognize complex sounds and filter out the meaningful information. In this way, useless noise is dropped and the raw information is distilled. Today, sensors can easily collect tons of raw audio data; however, processing them to get meaningful information remains arduous. Many researchers hope to design a human-like machine to alleviate this kind of problems, and one important and fundamental branch of it is called audio classification.

Currently, audio classification is used to distinguish audio samples by key words, intonation and accent. Audio classification can lead to real time transcription and translation of audio. The majority of audio classification research focuses on a specific classification task to obtain high accuracy. However, due to the complexity of audio data, various techniques must be employed to analyse the data.

In order to effectively classify the audio data, the features must be extracted from the audio sample. Three widely used techniques for audio classification research are Mel-Frequency Cepstral Coefficient (MFCC), Zero-Crossing Rate (ZCR) and Linear Predictive Coding (LPC)~\citep{qawaqneh2017deep, poria2015towards}. MFCCs have been used for feature extraction to improve speaker recognition. After this, Support Vector Machine (SVM) and Gaussian Mixture Model (GMM) are applied to do the classifications~\citep{pedersen2007accent, shegokar2016continuous}.

Recently, Deep neural networks (DNNs) and more specifically Convolutional neural networks (CNNs) have been used to automatically learn feature representations from complex data~\citep{yi2010foundations}. This universal technique has been applied in many areas to replace ad hoc function designs and has shorten a decade-long development period to a few months. The related applications have been seen in the audio classification area. For example, DNNs in conjunction with transformed MFCCs have been used to improve the accuracy of speaker age classification~\citep{qawaqneh2017deep}. Other researchers have used DNNs for cepstral feature extraction of audio samples~\citep{fu2013optimizing}. CNNs are able to deal with complex nonlinear mappings and can share weights across the input, which allows for translation invariance of the input. 

Most of the DNN-based algorithms need to convert the original audios into spectrograms before processing them. Spectrograms provide a visual representation of the frequencies with respect to time. Methods that use a time distributed approach~\citep{espi2015exploiting, lim2016speech} split the spectrogram into frames to create a time-distributed spectrogram. The time-distributed spectrogram is used as the input into the CNN to train the model to distinguish local features at different time steps. A different approach~\citep{leng2016employing} to audio classification splits the spectrogram along frequency to create a frequency-distributed spectrogram. Using this approach allows for the model to learn features based on various frequencies. 

Although the models based on spectrograms have achieved great successes, there are some intrinsic problems that are hard to eliminate. In particular, the function to generate spectrograms is independent from the later classification process. Practitioners must generate spectrograms from the audios before training the models. As a result, the spectrogram-generating function cannot be jointly optimized with the classification networks, which would considerably harm the performances of the algorithms. Besides, the spectrogram-generating process spans the originally one-dimensional audio data into three dimensions (one for time, one for frequency and one for three color channels: red, green and red), which makes the representation sparse (thus hard to learn) and adds extra noises that could interfere the later classification process. 

In this paper, we propose a new audio classification algorithm with an attention mechanism for the selection of the audio classifiers. We name the algorithm Classifier-Attention-Based CNN (CAB-CNN). Compared to the other DNN-based algorithms, 
\begin{enumerate}
    \item unlike the attention-based algorithm proposed by Wu et al. \citep{wu2018audio}, our attention unit dynamically assigns importance weights to a list of classifiers rather than attend to different frequencies and time intervals. This design let a single classifier only need to focus on a small portion of features. So, a classifier only needs to possess a small model capacity and does not need to have a large number of capacity. Therefore, the classifiers are much easier to be trained.
    \item since every single classifier only needs to learn a simple feature for distinguishing accents in principle. They can be trained easily and fast. Therefore,  the CAB-CNN model is more robust and have more stable performance in the independent training and testing processes. 
\end{enumerate}

We test the CAB-CNN model using UT-Podcast corpus \citep{HANSEN201619} by implementing an accent classification task. The test results corroborate what we have just claimed. Compared to the state-of-the-art algorithms~\citep{simonyan2014very, wu2018audio}, the CAB-CNN model has over $10\%$ improvements on all test scores and has reached $95.99\%$ test accuracy.

The rest of the paper is organized as follows. In Section~\ref{sec:relatedWork}, we introduce more techniques that have been used in the audio classification problems. In Section~\ref{sec:approach}, we propose our new algorithm formally. In Section~\ref{sec:experiment}, we test our algorithm on UT-Podcast corpus and compare its performance with some popular and the state-of-the-art algorithms. We conclude the paper in Section~\ref{sec:conclusion} and finally, we discuss some potential future work in Section~\ref{sec:futurework}.

\section{Related Work}
\label{sec:relatedWork}
The audio classification algorithms can be generally divided into two parts: the feature extraction part and the classification part \citep{8682912}. The feature extraction parts are mostly implemented by CNNs as they can efficiently extract characteristics from raw data \citep{6854950, 7952601}. 

The implementations of CNNs can be grouped into two classes based on how they preprocess the input audios: waveforms \citep{6854950,Lee2018} or spectrograms~\citep{wu2018audio,7952601,Choi2016}. The waveform-based method process the input data as an 1D data array directly, while the spectrogram-based implementations have to convert the raw audio files into spectrograms by Fourier transform first. Compared to the waveform-based method, the spectrogram-based methods manually extract frequency informations and plot them as a heat map. In other words, the strengths of the frequencies at each moment are indicated by the color or brightness. This preprocessing may facilitate CNNs to find frequency related features. In comparison, the waveform-based algorithm processes the raw audio files directly without involving plotting any graphs which are likely to introduce extra noises and/or make the data structure sparse. Besides, the entire  algorithm (data extraction and classification parts) can then be trained together and tuned jointly while the spectrogram-based methods have no control upon the spectrogram plotting part. 

The Multi-task Learning (MTL) method \citep{Caruana1997} has been used for multiple audio classification tasks. MTL is a focus of machine learning in which multiple learning tasks are solved simultaneously, improving the accuracy of multiple classifications by narrowing the gap between the training and testing errors~\citep{Baxter1995}. By employing a shared hidden layer, neural networks can use the MTL method~\citep{Caruana1997}. Studies have shown that MTL-SVM based models have better performance than task-specific SVM models~\citep{zhang2015recognizing}. The reason why the MTL works is that those factors that explain the variation of data could be shared among various tasks.

Despite many audio classification techniques being effective for a specific classification class, researchers have used convolutional deep belief networks (CDBNs) to classify audio data with high performance over multiple audio classification tasks~\citep{lee2009unsupervised}. The use of DNNs for cepstral feature extraction has also been used for multiple audio classification tasks~\citep{fu2013optimizing}.

Researchers are using deep residual networks (ResNets) along with a gate mechanism in order to extract feature representation in audio data. This was shown to be more effective with multiple audio classification tasks and has achieved higher accuracy compared to task specific models that were trained separately \citep{Zeng2019}.

\begin{figure}[t!]
    \centering
    \includegraphics[width=0.4\columnwidth]{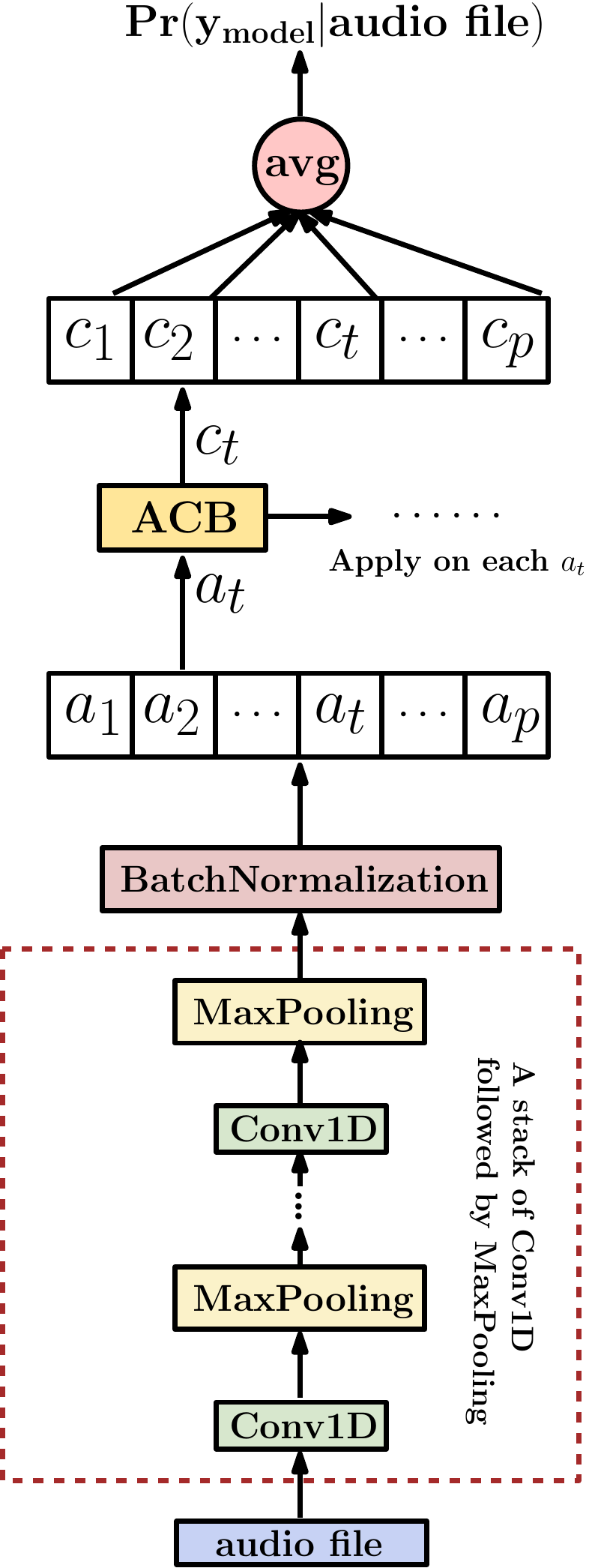}
    \caption{The complete architecture of CAB-CNN. The original audio file is first fed into a stack of CNNs and MaxPooling layers to get a ``distilled'' representation $a$. The batch normalization is applied at the end for making the model easier to train. For each time interval $t$, the ACB (the detailed design is presented in Fig~\ref{fig:attention_unit}) processes the representation of each $a_t$ and outputs the probability of the classes $c_t$. Finally, the output class probability is the unweighted average of $c_t$.}
    \label{fig:alg_arch}
\end{figure}
\begin{figure}[]
    \centering
    \includegraphics[width=0.75\columnwidth]{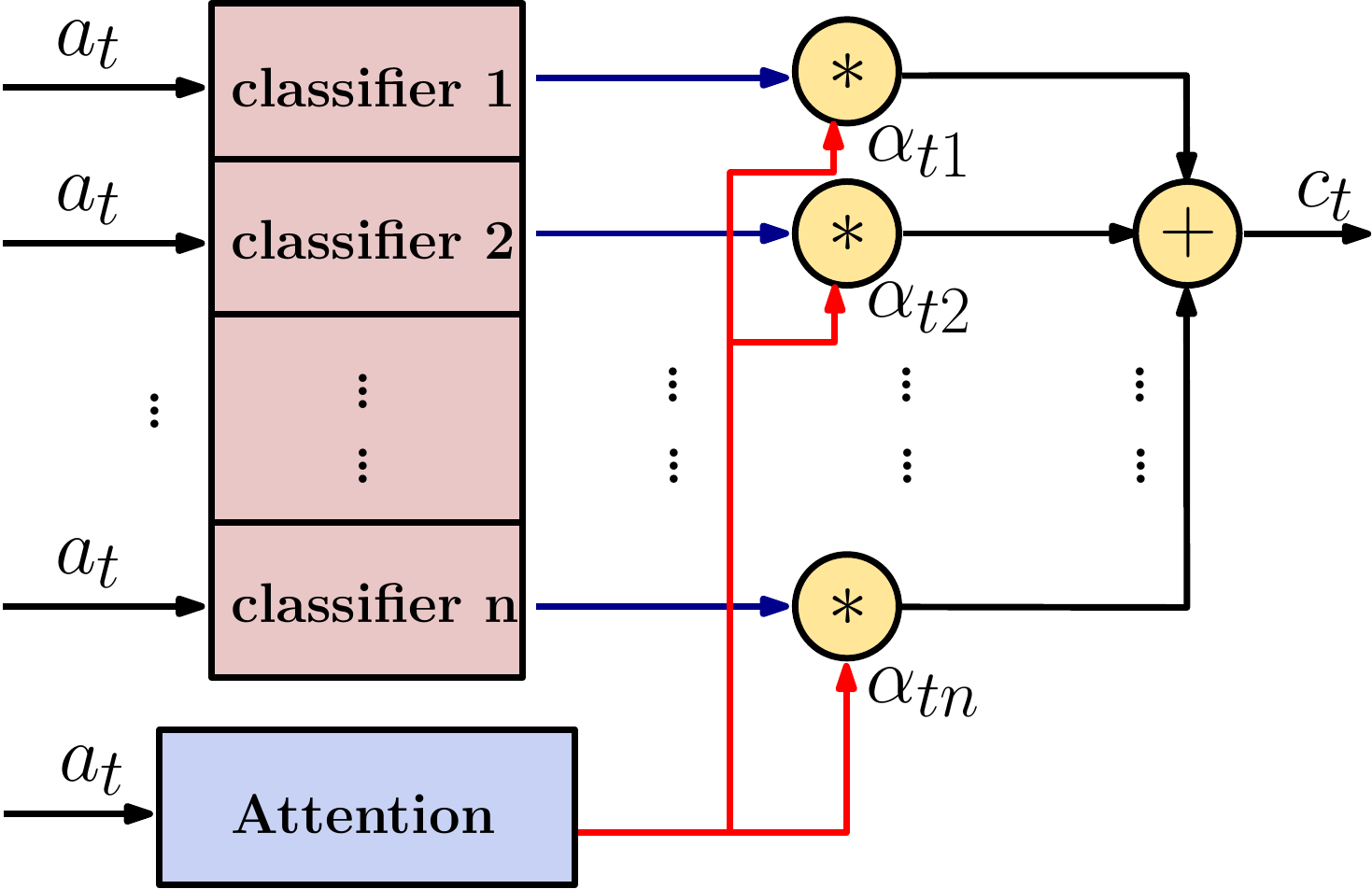}
    \caption{The attention-based classifier block (ACB) at time $t$. The block consists of an array of $n$ classification unit and an attention mechanism. The block receives an representation ($a_t$) of the audio file at time $t$ and then let each classifier implement an classification and the attention unit (attn) produce the importance weights $\alpha_{ti}$ for each classifier.
    The context vector $c_t$ then equals to the sum of the weighted outputs of the classifiers.}
    \label{fig:attention_unit}
\end{figure}

\section{Approach}
\label{sec:approach}
In this section, we introduce our classification algorithm with an attention mechanism for the selection of the audio classifiers. We name the algorithm Classifier-Attention-Based CNN (CAB-CNN), and its complete architecture is plotted in Fig~\ref{fig:alg_arch}.

\begin{figure}
	\centering
    \includegraphics[width=\columnwidth]{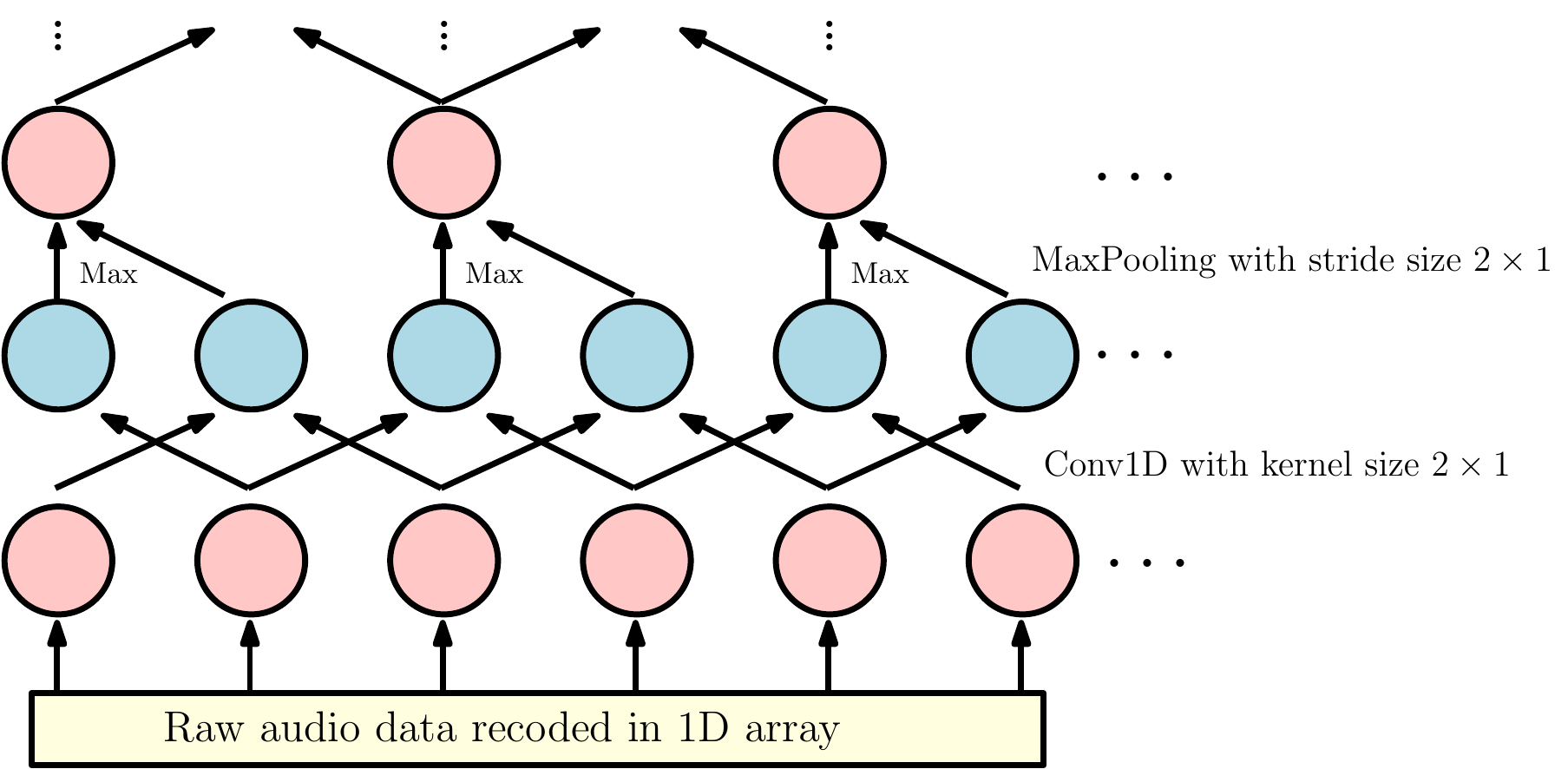}
    \caption{1D-CNN followed by a MaxPooling layer. The kernel of 1D-CNN can extract features from the raw data with very high parameter efficiency. The MaxPooling layer can then distill the data fed by the lower layers and thus reduces the output volume. In our algorithm, we have applied a stack of such structure to recursively distill information. In this way, the features that are useful for the later classification task are preserved, while those irrelevant informations are removed.}
    \label{fig:convInfoDistill}
\end{figure}

The key part of the CAB-CNN algorithm is the attention-based classifier block (ACB) containing $n$ classifiers and an attention unit (see Fig~\ref{fig:attention_unit}). In order to improve both training and statistical efficiency, we do not feed original audio to the block directly. Instead, we use a ``distilled'' representation $a$, which is generated by feeding the original data into a stack of 1D-CNN followed by MaxPooling layers (see Fig~\ref{fig:convInfoDistill}). We do so in order to 
\begin{enumerate}
    \item enhance local features,
    \item decrease the input data size of the classifier ACB, and
    \item preserve a one-to-one correspondence relationship (over time axis) between the original data and the generated representations.
\end{enumerate}

Roughly speaking, as CNN preserves the spatial information of the audio file, it partitions the original audio file into $p$ intervals (the value of $p$ depends on the architectures of CNNs, MaxPooling layers and the length of the input file) and applying the same transformation to extract features. Notice that these intervals would have some overlaps, which depends on the architectures of CNNs and MaxPooling layers. In Fig~\ref{fig:convInfoDistill}, we can observe that the first node of the MaxPooling layer covers the first three inputs of the raw data array while the second node covers the second to the fifth. Although the overlaps may cause some ambiguity of what the extracted features represent, they have no significant effect on the performance of the algorithm.

As our feature extraction algorithm preserves the spatial information, we can list its output feature vectors in the order of time:
\begin{equation}
    a = [a_1, a_2, \cdots, a_t, \cdots, a_p]
\end{equation}
where $a_t$ is the representation of the $t$-th time interval in the original data.

For each $a_t$ in the list, the ACB outputs a weighted average of the probability vector generated by the classifiers. 

In more details, suppose the ACB has $n$ classifiers, and there are $m$ classes to distinguish. Then, when receiving $a_t$,  $\text{classifier}$ i produces a classification probability vector $\mathbf{c}_{ti}\in \mathbb{R}^m$ for $i = 1,2, \cdots n$ and the attention unit generates the importance weights 
$$\alpha_t = [\alpha_{t1}, \alpha_{t2}, \cdots, \alpha_{tn}].$$
After this, the block outputs
$$c_t = \sum_{i=1}^{n}\alpha_{ti}\mathbf{c}_{ti}.$$

We can explain the design of the importance weights in two ways. First, they show how important a classifier is when classifying the audio at time $t$. Also, they represent the confidence that each classifier gives a correct output. 

By using this attention mechanism, we relieve a single classifier from having to distinguish a large set of features related to an audio classification task. Instead, one classifier only needs to focus on a certain type of features. 

In more details, assume that we need to classify some audios by the accents of the speakers. For a person, he/she may use the following strategy: 1) identify whether some certain features are present 2) if a feature is present and exclusively belongs to an accent, then we can say the audio is of this accent. Our algorithm works in a similar way. In particular, the attention unit identifies which features are present and lets the corresponding classifiers do the classification. This is done by assigning those classifiers with high importance weights. In this way, each classifier only needs to focus on a small subset of all available features, which thus can be trained easily and has a high predication accuracy. 

By feeding $a_t$ into the classifier ACB, we get a list of $c_t$ representing the predicted probabilities of the classes at time $t$. At last, we simply take the unweighted average over all $c_t$ to make the final predication. That is,
$$Pr(y_\text{model})= \sum_{t\geq 1}c_t.$$
Notice that this unweighted average implies a prior: the features implying the label of the classes have the identical probability of being active in each time interval. 
\section{Experiments}

In this section, we introduce our experimental mythology and the dataset for testing our new algorithm. We make some quantitative comparisons among the state-of-the art, a few popular neural network implementations and our new algorithm, CAB-CNN. The experimental results show that our new algorithm has a considerably better performance than the state-of-the-art. Moreover, we also provide more experimental results to show the behaviour of our algorithm.

We test our algorithm by performing accent classification task based on UT-Podcast corpus \citep{HANSEN201619}. This corpus contains audios of three English accents: American (US), Australian (AU) and Great British (GB). In the original dataset, the distributions of three accents in the training and test data are significantly different. To fix this problem, we mix them and take $60.0\%$, $10.0\%$ and $30.0\%$ of samples for training, validation and test. We detail the allocation of samples in Table~\ref{table:dataAllocation}.
\begin{table}[h!]
\caption{The allocation of samples of UT-Podcast corpus for training, validation and test.}\label{table:dataAllocation}
\begin{center}
	\begin{tabular}{|C{1.7cm}||c|c|c||c|}\hline
	 & {\bf US} & {\bf UK} & {\bf AU} & {\bf Total}\\\hline\hline
	{\bf Training}         &   387 &   347     &   458     &   1192\\\hline
	{\bf Validation}    &   65  &    58     &   76      &   199\\\hline
	{\bf Test}          &   194 &   174     &   230     &   598\\\hline\hline
	{\bf Total}         &   646 &   579     &   764     &   1989\\\hline
	\end{tabular}
\end{center}

\end{table}

\noindent
\textbf{Preprocessing of the audio file.} We need to preprocess the audio files to reduce their sizes and filter out noises to facilitate the learning of the model. In particular, we first normalize the audio by subtracting the mean followed by dividing by the standard deviation. Then for every second of the audio, we partition the audio into $4,000$ sub-intervals and pick the maximum element of each sub-interval to produce the input array. \newline

\label{sec:experiment}
\begin{table*}[]

\begin{center}
\caption{The layer configurations of CAB-CNN}\label{table:paraConfig}
\resizebox{1\textwidth}{!}{
	\begin{tabular}{p{2.9cm} p{6.8cm} p{2.1cm}}\hline
	\textbf{Layer Name}             & \textbf{Parameters}    & \textbf{Activation}    \\[0.6ex]
	Conv1D                  & \# filters: 16,  size: $4 \times 1$, padding: same        & ReLU\\[0.4ex]
	MaxPooling              & size: $4 \times 1$, stride: $2\times1$,  padding: same    & Dropout(0.15)\\[0.4ex]
	Conv1D                  & \# filters: 32,  size: $4 \times 1$, padding: same        & ReLU\\[0.4ex]
	MaxPooling              & size: $4 \times 1$, stride: $2\times1$,  padding: same    & Dropout(0.15)\\[0.4ex]
	Conv1D                  & \# filters: 32,  size: $10 \times 1$, padding: same       & ReLU\\[0.4ex]
	MaxPooling              & size: $10 \times 1$, stride: $5\times1$,  padding: same   & Dropout(0.1)\\[0.4ex]
	Conv1D                  & \# filters: 128, size: $10 \times 1$, padding: same       & ReLU\\[0.4ex]
	MaxPooling              & size: $10 \times 1$, stride: $5\times1$,  padding: same   & \\[0.4ex]
	BatchNormalization      & N/A  & N/A                              \\[0.4ex]
	ACB                     & $40$ Classifiers and $1$ attention block &   N/A   \\[0.4ex]
	Average                 &  N/A &   N/A \\\hline
	\end{tabular}
}
\end{center}
\end{table*}

\noindent
\textbf{Model configurations.} The detailed configurations of our model for testing are listed in Table~\ref{table:paraConfig}. For the configurations of the ACB, we simply use multilayer perceptrons to implement the classifiers and the attention block. In particular, all the classifiers consist of two fully connected layers of eight and four neurons with the ReLU activation functions, followed by a softmax layer to classify three accents. For the attention block, it has two fully connected layers of size $160$ and $80$ and a softmax layer for producing importance weights for $40$ classifiers. We also use the ReLU functions for introducing non-linearity. \newline

\noindent
\textbf{Model training.} We implement our model using the Keras library and train it from scratch by Adam optimizer \citep{arXiv:1412.6980} on NVIDIA GeForce GTX 1080 GPU. The loss is defined by the regular cross entropy function. We tested our model using at most the first $T$ seconds of the audio files, where $T = 5$s, $10$s, $30$s and $60$s (that is, if the audio length is less than $T$, we use the whole audio; otherwise, we only use the first $T$ seconds). In the following discussion, we add time length after CAB-CNN to specify which implementation we are referring to. We set the batch size to $128$. Since the Keras library requires that the input size of the neural network in a single batch must be the same, we truncate the sizes of the batch inputs just equal to the shortest one of that batch. We also apply the early stopping technique with patience $15$ to fight against the overfitting problem.\newline

\noindent
\textbf{Configurations of other algorithms for comparison.} We demonstrate the performance of CAB-CNN by comparing it with some typical CNN architectures and the state-of-the-art algorithm. In particular, they are GatedResNet~\citep{Zeng2019}, Alexnet~\citep{krizhevsky2012imagenet}, VGG11$_A$~\citep{simonyan2014very}, ResNet18~\citep{he2016deep} and AttentionCNN~\citep{wu2018audio}. All these implementations need to preprocess the audios by converting them into spectrograms. In our experiments, we simply use the sizes suggested by the authors. For GatedResNet and Alexnet, the audios are converted to the graph of size $256 \times 256$, and for VGG11$_A$ and ResNet18, the spectrograms have size $224 \times 224$. All these graphs have three channels: red, green and blue. Regarding AttentionCNN, the sprectograms are in grayscale and of size $256 \times 256$. For all these models, we apply the original configurations presented in the papers except modifying the softmax layer for fitting our tasks. All these models are trained from scratch with the stochastic gradient descent (SGD) optimizer with learning rates $0.001$, weight decay $10^{-8}$ and momentum $0.9$. The batch size is set to $48$. We also apply the early stopping technique with the patience equal to $15$.

\begin{table*}[t]
  \caption{The best result of each audio classification algorithm among eight parallel\protect\linebreak tests  over the remixed UT-Podcast corpus.}
  \begin{center}
    \label{table:algPerfoemance}
    \begin{tabular}{|C{2.8cm}|C{2cm}|C{2.8cm}|C{2cm}|}\hline
      \textbf{Algorithm} & \textbf{Accuracy} &  \textbf{Recall$_{\text{unweighted}}$} & \textbf{F1-Score}\\
	  \hline
      \hline
      GatedResNet~\citep{Zeng2019} & $0.7939$& $0.7487$ & $0.7557$\\
      \hline
      Alexnet~\citep{krizhevsky2012imagenet} & $0.6793$ & $0.6115$ & $0.6152$ \\
      \hline
      VGG11$_A$~\citep{simonyan2014very} & $0.8645$& $0.8244$ & $0.8331$\\
      \hline
      ResNet18~\citep{he2016deep} & $0.8015$& $0.7424$ & $0.7512$\\
      \hline
      AttentionCNN~\citep{wu2018audio} & $0.8626$ & $0.8257$ & $0.8330$\\\hline
      CAB-CNN-30s & $\bf{0.9599}$ & $\bf{0.9424}$ & $\bf{0.9523}$ \\
      \hline
     \end{tabular}
  \end{center}
\end{table*} 
\begin{table*}[t]
  \caption{Performances of the CAB-CNN using at most the first $T$ seconds\protect\linebreak of the audios ($T=5, 10, 30$ \& $60$).}
  \begin{center}
    \label{table:performanceCAB-CNN}
    \begin{tabular}{|C{2.5cm}|C{2cm}|C{2.8cm}|C{2cm}|}\hline
      \textbf{Algorithm} & \textbf{Accuracy} & \textbf{Recall$_{\text{unweighted}}$} & \textbf{F1-Score}\\
      \hline
      \hline
      CAB-CNN-5s & $0.9198$ & $0.8953$ & $0.9092$ \\
	  \hline
	  CAB-CNN-10s & $0.9542$ & $0.9383$ & $0.9475$ \\
      \hline
	  CAB-CNN-30s & $\bf{0.9599}$ &  $\bf{0.9424}$ & $\bf{0.9523}$ \\
	  \hline
	  CAB-CNN-60s & $0.9466$  & $0.9228$ & $0.9356$ \\
	  \hline
    \end{tabular}
  \end{center}
\end{table*}

Table~\ref{table:algPerfoemance} lists the accuracy, unweighted recall, F1-Score of the selected algorithms by training and testing them on the remixed UT-Podcast corpus. The highest value of each score is bold. For each algorithm, we repeat the experiment for eight times and list the result having the highest accuracy. Since the CAB-CNN has the best performance when processing at most the first 30 seconds of the audio file, we only list the scores of CAB-CNN-30s here. We summarize the performances of CAB-CNN with other configurations in Table~\ref{table:performanceCAB-CNN}.

Suppose the test dataset contains $N$ samples. Let $y^{(i)}_{data}$ denote the actual accent of the $i$-th sample in the test data and $y^{(i)}_{model}$ the predicted result generated by the algorithms. Then the scores are calculated by
$$
    \textit{Accuracy} = \frac{\sum_{i=1}^N\mathbf{1}_{\left(y^{(i)}_{data}=y^{(i)}_{model}\right)}}{\textit{N}}.
$$

Let $L$ denote the set of the accents for classifying and $T$ the test sample set. For accent $l\in L$, $T_l$ denotes the samples having accent $l$ in $T$. Let $|S|$ be the size of set $S$. Then we define 
$$\textit{Recall}_{\textit{unweighted}} = \frac{1}{|L|}\sum_{l\in L}\textit{Recall}(l),$$
where 
$$\textit{Recall}(l) = \frac{\sum_{i\in T_l}\mathbf{1}_{\left(y^{(i)}_{data}=y^{(i)}_{model}\right)}}{|T_l|}.$$
Let 
$${\textit{Precision}(l) = \frac{\sum_{i\in T_l}\mathbf{1}_{\left(y^{(i)}_{data}=y^{(i)}_{model}\right)}}{\sum_{i\in T}\mathbf{1}_{\left(y^{(i)}_{model}=l\right)}}.}$$ 
Then 
$$\textit{F1-Score}=\frac{1}{|L|}\sum_{l\in L}\textit{F1-Score}(l),$$
where 
$$\textit{F1-Score}(l) = 2\cdot \frac{\textit{Precision}(l)\cdot \textit{Recall}(l)}{\textit{Precision}(l) + \textit{Recall}(l)}.$$
\begin{table*}[t!]

\caption{The confusion matrix of the best results of $\text{VGG}11_A$ and AttentionCNN tested on the UT-Podcast dataset.}\label{table:ConfMatrixVGG11andAttentionCNN}
\begin{center}

\begin{tabular}{|l||c|c|c||c|c|c|}
\hline
& \multicolumn{3}{c||}{$\text{VGG}11_A$} & \multicolumn{3}{c||}{AttentionCNN} \\ 
\cline{1-4}\cline{5-7}
\backslashbox{Act.}{Pred.}  
& {\textbf{AU}}  & {\textbf{US}}  & {\textbf{UK}}     & {\textbf{AU}}  & {\textbf{US}}  & {\textbf{UK}}\\ 
\cline{1-4}\cline{5-7}
{\textbf{AU}} & 213 & 7 & 10  & 208 & 11   & 11 \\ 
\cline{1-4}\cline{5-7}
{\textbf{US}} & 9   & 176 & 9  &  7   & 179 & 8 \\ 
\cline{1-4}\cline{5-7}
{\textbf{UK}} & 24  & 12  & 64 & 22   & 13 & 65\\
\hline
\end{tabular}

\end{center}
\end{table*}

\begin{table*}[t!]
\caption{The confusion matrix of the CAB-CNN using at most the first $T$ seconds\protect\linebreak of the audios ($T=5, 10, 30$ \& $60$).}\label{table:ConfMatrixCAB}
\begin{center}
\vspace{-1.5em}
\resizebox{1\textwidth}{!}{%
\begin{tabular}{|l||c|c|c||c|c|c||c|c|c||c|c|c||}
\hline
& \multicolumn{3}{c||}{CAB-CNN-5s} & \multicolumn{3}{c||}{CAB-CNN-10s} &  \multicolumn{3}{c||}{CAB-CNN-30s} &  \multicolumn{3}{c||}{CAB-CNN-60s}  \\ 
\cline{1-4}\cline{5-7}\cline{8-10}\cline{11-13}
\backslashbox{Act.}{Pred.}  
& {\textbf{AU}}  & {\textbf{US}}  & {\textbf{UK}}     & {\textbf{AU}}  & {\textbf{US}}  & {\textbf{UK}} &{\textbf{AU}}  & {\textbf{US}}  & {\textbf{UK}}                   &  {\textbf{AU}}  & {\textbf{US}}  & {\textbf{UK}}                    \\ 
\cline{1-4}\cline{5-7}\cline{8-10}\cline{11-13}
{\textbf{AU}} & 219 & 9 & 2  & 222 & 8   & 0 & 226 & 4   & 0 &  224 & 5 & 1 \\ 
\cline{1-4}\cline{5-7}\cline{8-10}\cline{11-13}
{\textbf{US}} & 9   & 185 & 0  &  1   & 192 & 1 &  2   & 191 & 1 &   3   & 191 & 0\\ 
\cline{1-4}\cline{5-7}\cline{8-10}\cline{11-13}
{\textbf{UK}} & 10  & 12  & 78 & 7   & 7 & 86 & 8   & 6   & 86  & 10  & 9   & 81\\
\hline
\end{tabular}
}
\end{center}
\end{table*}
From Table~\ref{table:algPerfoemance}, we can observe that GAB-CNN-30s has a significant better result than other models. In particular, compared with the state-of-the-art algorithms VGG11$_A$~\citep{simonyan2014very} and AttentionCNN~\citep{wu2018audio}, our algorithm has more than $10$ percent improvement in accuracy, recall and F1-Score.  

By comparing the test results of the GAB-CNN with various maximum length of input audio (see Table~\ref{table:performanceCAB-CNN}), we can observe that GAB-CNN has the best performance when the maximum length of the audio is set to be $30$s. It is quite reasonable that the algorithm could suffer from a low performance if the input audio length is too short. In more details, a short audio may not contain enough features for accent classification. Consider the extreme case that the input audio only contains one word. Then if the pronunciation of this word is the same among all three accents, there is no way to classify it and the classification output will be randomly picked. 

A decline of the performance can be also observed if the input audio is too long. This decline could be partially caused by the avg layer that summarizes the classification results in each time interval and outputs the predication in probability distribution (see Fig~\ref{fig:alg_arch}). As what we have mentioned in the previous paragraph, it is possible that some words do not contain information for the accent classification. When a long audio is given, this kind words may become prevalent. Then the random results corresponding to these words would weaken or even conceal the true results that are generated from those classifiable words.

We list the confusions matrices of models VGG11$_A$, AttentionCNN and GAB-CNN with difference configurations in Table~\ref{table:ConfMatrixVGG11andAttentionCNN} and Table~\ref{table:ConfMatrixCAB}. We can obverse that the score improvements are contributed by all kinds of classification problems. For instance, from Table~\ref{table:ConfMatrixVGG11andAttentionCNN}, we can see that both VGG11$_A$ and AttentionCNN models are likely to misclassify an UK accent as AU one, while GAB-CNN-30s can reduce this kind of error by two thirds.


\begin{table*}
\caption{The average algorithm performances by implementing eight parallel \protect\linebreak tests over the remixed UT-Podcast corpus.}\label{table:avgPerformance}
\begin{center}
\vspace{-2em}
\begin{tabular}{|C{3.4cm}|C{2cm}|C{3cm}|C{2cm}|} 
\hline
\backslashbox{\textbf{Model}}{\textbf{Metrics}}  & \textbf{Accuracy} & \textbf{Recall$_{\text{unweighted}}$}  & \textbf{F1-Score}\\ 
\hline\hline
GatedResNet     & $0.6594$ & $0.5854$  & $0.5686$\\ 
\hline
Alexnet         &  $0.5795$  &  $0.5590$  & $0.4793$\\ 
\hline
VGG11$_A$          &  $0.6527$ &  $0.6227$ & $0.6330$ \\ 
\hline
ResNet18        &  $0.7950$  & $0.7373$ & $0.6908$ \\ 
\hline
AttentionCNN    &  $0.7410$ &  $0.6012$  & $0.4794$ \\ 
\hline
CAB-CNN-30s     & $\bf{0.9370}$ &  $\bf{0.9163}$  & $\bf{0.9282}$ \\
\hline
\end{tabular}
\end{center}

\end{table*}

Another highlight of our model is that the GAB-CNN is more robust. Specifically, while the other models require repeating the training processes for many times before finding an acceptable parameter solution, our model can always converge to a solution having good test scores. We justify this statement by averaging the scores of each algorithm over the eight parallel tests. We summarize the results in Table~\ref{table:avgPerformance}. From the table, we can observe that there is not a large gap between the best and the average performances of the GAB-CNN. In comparison, large drops can be observed for other algorithms. This observation implies that, the attention mechanism integrated in our model makes it easy for the model to be trained. As we have discussed in Section~\ref{sec:approach}, the attention block relieves a classifier from having to identify all features that are different among various accents. Instead, it is enough for a classifier to only focus on a list of similar features. In this way, a classifier does not have to possess a large model capacity which requires a huge number of parameters. Therefore, both training and statistical efficiencies are improved, and the model is much easier to be trained.

\section{Conclusion}
\label{sec:conclusion}
In this paper, we have proposed an attention-based audio classification deep neural network named the Classifier-Attention-Based CNN (CAB-CNN). Unlike many of other DNN implementations in this area, our algorithm does not need to convert the audio files into spectrograms as a preprocessing step and thus avoids the unnecessary introduction of noises and makes jointly training on the whole model possible. Unlike Wu et al.'s attention-based model that attends to different frequencies and time intervals \citep{wu2018audio}, our model instead uses the attention block to select the proper classifiers to distinguish the input audios. This design makes our algorithm more robust and it has a significant better performance than all published neural network models in this area. Our work shows that an accent classification algorithm can gain a remarkable performance improvement by deploying a list of simple and specialized classifiers with an attention mechanism determining which classifier's result is more trustworthy and thus has a larger portion in the final predication. 

We tested our model by performing accent classification tasks on the UT-Podcast corpus. Compared to the state-of-the-art algorithms~\citep{wu2018audio,simonyan2014very}, our model has more than $10\%$ improvements on all test scores and has reached $95.99\%$ test accuracy.

\section{Future work}\label{sec:futurework}
Although, in this paper, we have presented an algorithm with a new architecture that has a descent improvement compared to the state of the art, its performance still has large room to be improved. 

In our implementation, we simply use fully connected layers to implement the classifiers and the attention mechanism. We believe that some dedicated implementations can further improve its performance. Besides, as we have mentioned in Section~\ref{sec:experiment}, our algorithm suffers from a performance decline if the input audio length is too long. To fix this problem, we could design another learning algorithm to find the best length of input audio. Also, we could train an algorithm to locate the words that are accent classifiable and only input those words into our classification network.

\bibliography{bibliography}

\end{document}